# Out-of-plane emission of trions in monolayer WSe$_2$ revealed by whispering gallery modes of dielectric microresonators

*Daniel Andres-Penares[1,†], Mojtaba Karimi Habil[2,3], Alejandro Molina-Sánchez[1], Carlos J. Zapata-Rodríguez[2], Juan P. Martínez-Pastor[1,4], and Juan F. Sánchez-Royo[1,4*]*

[1]   ICMUV, Instituto de Ciencia de Materiales, Universidad de Valencia, P.O. Box 22085, 46071 Valencia, Spain

[2]   Department of Optics and Optometry and Vision Sciences, University of Valencia, c/ Dr. Moliner 50, Burjassot 46100, Spain

[3]   Faculty of Physics, University of Tabriz, 51664 Tabriz, Iran

[4]   MATINÉE: CSIC Associated Unit-(ICMM-ICMUV of the University of Valencia), Universidad de Valencia, P.O. Box 22085, 46071 Valencia, Spain

The manipulation of light emitted by two-dimensional semiconductors grounds forthcoming technologies in the field of on-chip communications. However, these technologies require from the so elusive out-of-plane photon sources to achieve an efficient coupling of radiated light into planar devices. Here we propose a versatile spectroscopic method that enables the identification of the out-of-plane component of dipoles. The method is based on the selective coupling of light emitted by in-plane and out-of-plane dipoles to the whispering gallery modes of spherical dielectric microresonators, in close contact to them. We have applied this method to demonstrate the existence of dipoles with an out-of-plane orientation in monolayer $WSe_2$ at room temperature. Micro-photoluminescent measurements, numerical simulations based on finite element methods, and *ab-initio* calculations have identified trions as the source responsible for this out-of-plane emission, opening new routes for realizing on-chip integrated systems with applications in information processing and quantum communications.

1. **Introduction**

The race to develop high-performance photonic and optoelectronic devices which take advantage of the distinctive properties and versatility of two-dimensional (2D) semiconductors[1–3] has already given rise, for example, to high-speed and high-responsivity waveguide-integrated photodetectors,[4] plasmonic nanocavities,[5–7] and optical resonators with enhanced light-matter interactions,[8–10] as the basis to construct nanolasers operating at room temperature.[11,12] Most of the optoelectronic and photonic devices developed so far are based on 2D transition metal dichalcogenide (TMD) semiconductors such as $MoS_2$, $MoSe_2$, $WS_2$ and $WSe_2$. Consequently, the characteristics of the performance of these devices strongly depend on the in-plane (IP) dipolar nature of the robust free excitons of the 2D TMDs.[13–15] Light from IP excitons can be easily extracted in stacked (vertical) devices. However, some emerging applications in the generation of radially polarized light in 2D TMDs[16] and integrated photonic chips can be efficiently enabled when out-of-plane (OP) excitons mediate light–matter interactions.[17–24] Although rare, OP excitons can be found in 2D semiconductors. For instance, atomically thin layers of Indium Selenide (InSe) have demonstrated, as 2D TMDs, potential applications for next generation electronics and optoelectronics[25–27] due to their highly tuneable band gap[28–30] and high electron mobility.[31] Nevertheless, unlike 2D TMDs, 2D layers of InSe have revealed to sustain luminescent free excitons with an intrinsic OP orientation.[14] This panorama puts into evidence the necessity of identify, among the different excitonic complexes existing in 2D semiconductors, those with a suitable dipole orientation for their optimal application in the emerging field of integrated photonic circuits based on 2D materials.[32–37] This question is particularly relevant for the development of novel devices based on monolayer (ML) TMDs, since strong spin-orbit coupling makes these materials offer, apart from bright and long-lived dark excitons,[38,39] high-order charge-complexes with unexplored dipolar characteristics, such as bound excitons, trions,[40–44] and biexcitons.[45] These complexes are technologically promising as they can manifest even at room temperature,[46–48] have an intrinsic charge and spin degrees of freedom that facilitate their manipulation by the application of an electric or a magnetic field,[45,49] possess a valley degree of freedom (as free-excitons), permit new optical gain mechanisms at extremely low carrier densities,[50] and can be further used as entangled photon sources.[51–55]

Back focal plane imaging is usually considered the method-of-choice to discern the orientation of the dipole responsible for the luminescent signal emitted.[13,14,56] However, it is challenging to use this technique to discriminate contributions from different sources to the luminescent signal that may eventually have a different dipolar orientation. To overcome this limitation, more sophisticated methods have been proposed, but their application is rather limited as they require an efficient excitation of surface plasmon polaritons.[57] In this work, we propose a versatile and simple method to spectroscopically elucidate the existence of an OP component in the photoluminescent response of a dipole which, at the same time, puts into evidence its feasibility to promote light coupling to a planar device. The proposed method relies on the different selective ability of IP and OP dipolar emission to excite whispering gallery modes of $SiO_2$ microspheres deposited on top of the emitting layers. We have applied this method to study the orientation of dipoles related to the different light contributions observed in ML $WSe_2$ at room temperature, these from trions and excitons. By microphotoluminescence (µ-PL) measurements, numerical simulations based on the finite element method (FEM), and *ab-initio* calculations of the excitonic states we have found a non-negligible OP dipolar component of trions in ML $WSe_2$, which contrasts to the well-established IP nature of its free bright excitons. These results establish that exciton complexes are excellent candidates for light manipulation in planar optoelectronic devices and photonic chips.

## 2. Results and discussion

Figure 1 puts into evidence pronounced lensing and strong light-coupling effects when, as is illustrated in Figure 1a, dielectric microspheres are deposited on top of 2D semiconductors. Figure 1b shows µ-PL spectra acquired in a bulklike (16 nm thick) InSe nanosheet deposited on a $SiO_2$/Si substrate and partly covered with $SiO_2$ microspheres (see optical image at the inset of Figure 1b). The µ-PL spectrum measured in a bare region of the InSe nanosheet shows an emission peak located at 1.24 eV, corresponding to the radiative free-exciton recombination processes occurring at energies reported for bulk InSe.[14,30] Although it is almost negligible in this spectrum, it is worth mentioning that a weak broad emission is detected at energies higher than 1.45 eV, which stems from the amorphous $SiO_2$ substrate. Compared to the µ-PL spectrum measured in the bare InSe nanosheet, the one acquired in a point of sample covered with $SiO_2$ microspheres shows a strong enhancement of the signal intensity (see Figure 1b and the μ-PL map in its inset) that makes visible even the weak luminescence coming from the $SiO_2$ substrate. This lensing effect, promoted by the microspheres, is not restricted to bulklike InSe nanosheets. In fact, the use of dielectric microspheres seems to provide for a tool to enhance the usually low-intensity PL signal of atomically-thin InSe nanosheets without damaging the sample, as it has been demonstrated for a 2D InSe nanosheet 6.5 nm-thick with an noticeable blueshift of its PL signal due to quantum-confinement effects (See Supplementary Figure S1).[30]

A similar lensing effect is observed in other 2D semiconductors. Figure 1c shows the μ-PL spectra acquired in a ML of $WSe_2$ deposited on a $SiO_2$/Si substrate, on which a $SiO_2$ microsphere has been placed (see the optical image at the inset of Figure 1c). In the bare region of the ML of $WSe_2$, the PL spectrum shows a double-peak structure whose deconvolution has been performed by assuming gaussian-like components.[48,58] In this way, deconvolution processes have allowed to resolve a main PL peak centered at 1.666 eV, coming from optical recombination of the $X^0$ neutral exciton, and a second and broad PL peak centered at 1.640 eV. Such low-energy feature has been usually attributed to recombination of trions, whose observation is hard at room temperature but possible in doped samples (as those used in this work) due to the higher probability of forming charged excitons in these samples than in intrinsic ones.[48,57,59,60] In line with that observed for InSe nanosheets, the μ-PL spectrum measured in the region of the $WSe_2$ nanosheet under the $SiO_2$ microsphere (see also the PL map inserted in Fig. 1c) shows an enhanced intensity signal, compared to that of the bare ML.

Clearly, the lensing effect described above is promoted by the difference between air and the microsphere refractive index,[61–63] increasing the excitation power density on the sample (Figure 2a, top panel) and consequently enhancing the global PL signal intensity. The lensing effect is not only restricted to the excitation process but extends also to the light-collection process (Figure 2a, bottom panel). Importantly, this effect tends to enhance the collection of light within the numerical aperture of our optical objective that would be eventually lost in the absence of the microsphere, as the far-field emission from OP dipoles.[64] Besides the lensing effect, the presence of the microspheres also promotes the appearance of strong resonances in the PL spectrum (see Fig. 1 and Supplementary Figure S1) which reveal a strong coupling of emitted light from dipoles existing under the dielectric microspheres to modes of the spherical microresonator (Figure 2b and 2c). The spectral response of these resonances, called whispering gallery modes (WGMs),[65–67] mostly depends on the precise diameter of the microsphere (Supplementary Figure S2), which may be used to fine-tune the relative weight of desired energy regions of the PL spectrum with respect to others. However, we would like to focus here on a photonic application of these microspheres on nanomaterials (See Supplementary Figures S3 and S4), connected to the ability of WGMs of dielectric microspheres to provide for a tool to discriminate the OP component, from the IP one, of dipoles underneath (i.e., the luminescent dipoles of the 2D nanosheet), as illustrated in Figure 2c.

Figure 2d shows the μ-PL spectra measured in the $SiO_2$/Si substrate, covering the 1.43-1.70 eV energy region where a weak emission from the amorphous $SiO_2$ overlayer has been detected (Figure 1b). One of these spectra has been acquired in the bare $SiO_2$/Si substrate (black continuous line), whereas the other has been recorded in a point of the substrate covered by a $SiO_2$ microsphere (red continuous line). In this second PL spectrum, the emitted light appears to couple to the WGMs of the microsphere on top, which gives rise to resonances that consist in relatively intense peaks separated to each other every 60-70 meV. Besides, the spectra reveal the additional presence of weak but clear resonances (see, for instance, the one at 1.675 eV) in between the more intense lines. To understand the origin of this pattern of alternating intense and weak resonances, we have calculated the spectral response of a dipolar emission when coupled to the WGMs of a spherical resonator of 4.66 µm in diameter, a value close to the nominal diameter of the microspheres (~5 µm, Methods section) that allows to nicely reproduce the experimental data. These numerical simulations are based on the treatment developed by Chew,[68] which disregards the effects of the substrate. See Supplementary Figure S3 for further details about the effects of a realistic environment in the coupled emitter-resonator, as obtained by FEM-based simulations. If we first consider that the dipole is oriented parallel to the z-axis -or OP- (Figure 2c), its radiated field is expected to couple to a discrete set of transverse magnetic (TM) modes of the spherical resonator (Figure 2d, orange curve), where the modal electric field is given in terms of the spherical vector wave functions $\textbf{\textit{N}}_{lm}^{(1)}$, with polar mode number $l = 1,2,...$ and azimuthal mode number $m = 0, \pm 1, ..., \pm l$.[69] Note that the magnetic field of the vertical-dipole radiation is azimuthally oriented parallel to $\hat{\boldsymbol{\varphi}}$, allowing the electric field of the scattered signal set in terms of the spherical vector wave functions $\textbf{\textit{N}}_{lm}^{(3)}$.[70] Peak resonances of the scattered field, labeled as $TM_{l,n}$ in Figure 2d (orange curve), are characterized by the solutions to the so-called modal equation and are classified by a radial mode number $n = 1,2,...$. They are thus characteristic from each *l*th-order TM mode, taking into account a generalized $2l + 1$ mode degeneracy when varying the azimuthal index $m$.[68] Nevertheless, due to the location of the electric dipole set along the z-axis, the highly-symmetric scattered fields lead to excitation of only one azimuthal mode $m = 0$ for each polar mode index. In addition, due to the relatively low $Q$ factors associated with the dielectric microspheres, one can only observe peaks in Figure 2d corresponding to the radial mode $n = 1$ (those with *l* between 20 and 23).

When tilting the dipole $\boldsymbol{p}$ from the OP configuration to have an IP dipolar component, excitation of transverse electric (TE) modes in the dielectric resonator starts to be allowed (see Supplementary Information file). Now, the electric field in the resonator include terms depending on the spherical vector wave functions $\boldsymbol{M}_{lm}^{(1)}$. As a result, a new set of natural frequencies associated with the peaks labeled as $TE_{l,n}$ will also appear in the spectrum of the scattered electric field accompanied by their corresponding $TM_{l,n}$ ones, involving the spherical vector wave functions $\boldsymbol{M}_{lm}^{(3)}$. The relative coupling strength of TM and TE modes depends on the tilting angle of the dipole. If the dipole is fully IP oriented (Figure 2c), the coupling strength to $TE_{l,n}$ modes is significantly higher than to their corresponding $TM_{l,n}$ ones (Figure 2d, purple curve). However, in comparison with the TM modes excited by the OP dipoles, both TM and TE modes excited by IP ones are less intense (Figure 2d). In this case, note that the excited azimuthal modes with TE and TM polarization are now $m = \pm 1$ exclusively.[68,70] On the contrary, the radial index $n = 1$ of the observed peaks is maintained unaltered here.

The nice match observed between experimental and simulation results summarized in Figure 2d reveals a clear preference of WGMs in dielectric microspheres to couple to light emitted from OP dipoles in detriment of that from IP ones. Naturally, this selective behaviour of WGMs provides for an easy, fast, and powerful method to detect the presence of OP dipoles in an emitting 2D layer. Furthermore, the observation of WGMs at a particular energy of the PL spectrum can be considered as a fingerprint of the ability of the emitted photons to propagate along a planar structure, somehow represented by a dielectric microsphere. Since each kind of radiative excitonic complex (free-exciton, trion, bi-exciton, etc) can be identified by its transition energy, the spectral distribution of WGMs in the PL spectrum would allow to identify and select, among all excitonic complexes of a system, the optimal candidates to be implemented in optoelectronic devices designed to operate in a particular planar configuration.

The method proposed above, to disentangle the OP component of radiative excitonic complexes and detect their potential ability to couple its light to horizontal devices, has been applied to 2D materials with promising applications in optoelectronics, photonics, and quantum communications. InSe and WSe$_2$ can be considered as paradigm systems which give rise to 2D semiconductors and devices with a completely different orientation of their respective free neutral excitons. Luminescent free excitons of 2D InSe have been recently demonstrated to be OP whereas those of ML WSe$_2$ are IP.[13–15] With the aim to emphasize the effects of the microspheres on the PL response of these kind of nanosheets, we show, in the middle panels of Figures 3a and 3b, the ratio between the μ-PL spectra acquired in a point of the nanosheet under the microspheres and in another of the bare one, for the 16 nm-thick InSe nanosheet and the ML WSe$_2$ already described in Figure 1, respectively. For the sake of clarity, top panels of Figure 3a and 3b show again the original μ-PL spectra acquired in each one of these nanosheets and bottom panels display the spectral response calculated for IP and OP dipolar emission eventually coupling to the WGMs of a spherical resonator of 4.69 μm in diameter, within the energy range enclosed by the PL response of each nanosheet, respectively. Results summarized in Figure 3 reveal a different behavior of the excitation of WGMs, depending on the nanosheet probed. The PL intensity ratio obtained for the 16 nm thick InSe nanosheet (Figure 3a, middle panel) shows an approximately constant enhancement factor of 4-5, in the energy range comprising the PL peak (1.20-1.35 eV), due to the lensing effect. Overlapped with this basement line, the spectrum shows clear and pronounced resonances at the energy positions expected for light coupling to the WGMs of the microspheres when the emitter is, precisely, an OP dipole (Figure 3a, bottom panel). A similar behaviour arises for thinner InSe nanosheets (Figure S1), although the microsphere used in this experience was slightly bigger (of 4.80 μm in diameter). These results support previous works reporting the OP orientation of radiative free excitons in InSe,[14] and allow us to conclude, by the relatively simple method used here, that radiative recombination processes of OP free excitons conform the whole room-temperature PL spectrum of InSe nanosheets. The dominant OP orientation of dipoles responsible for the room-temperature PL signal of InSe contrasts to these observed for ML WSe$_2$ (Figure 3b). Unlike InSe, the magnification of the PL signal promoted by the microspheres as well as the excitation of WGMs resonances tend to concentrate in the low-energy side of the PL spectrum of ML WSe$_2$ (Figure 3b, middle panel), being both effects strongly reduced at the energies of the main PL peak. In fact, the excitation of WGMs becomes particularly strong at light-emission energies as low as ~1.50 eV (see inset at the top panel of Figure 3b), far beyond the energy range usually observed for radiative recombination of free excitons in ML WSe$_2$ at room temperature (Figure 1c). Therefore, as occurs for InSe, WGMs resonances in the WSe$_2$ ML appear to be excited by its OP dipoles (Figure 3b, bottom panel). These results reflect that, as reported, luminescent free-excitons of ML WSe$_2$ are strongly IP.[13–15] However, light emitted from the recombination of trions, which notably contribute to the low-energy side of the PL spectrum of ML WSe$_2$ (Figure 1c), has a relevant OP dipolar component that promotes its coupling to the WGMs of the microspheres

The electronic states at the bandgap of ML WSe$_2$ are majorly tungsten d-orbitals and the spin-orbit interaction is strong, resulting in optical selection rules with IP and OP emission dipoles of distinct intensities (see the Supplementary Information for discussion of the optical selection rules).[38,55] In order to quantify the luminescent dipoles in monolayer WSe$_2$ we have calculated the exciton states for IP and OP polarized light. Figures 4a-b show the band structure of monolayer WSe$_2$ calculated along the M-K-Γ high-symmetry directions, indicating the electronic transitions relevant for the absorption of IP and OP light and the optical absorption including excitonic effects, respectively. The dark $X_z^0$ and bright $X^0$ exciton transition energies are marked with blue and red vertical lines, separated by 69 meV. In contrast to InSe, the optical selection rules have the origin in the spin-orbit interaction.[71,72] The strong effect of the spin-orbit interaction is evidenced by looking at the wave functions of excitons $X_z^0$ and $X^0$ in Fig. 4d. The excitonic wave function depends on the electron and hole position. In these calculations, we fix the hole position near the W atom and the resulting electronic density is displayed, with the $X_z^0$ excitonic configuration showing a higher electron density on the Se atoms that is responsible for its OP orientation component. Moreover, each exciton $X^0$ and $X_z^0$ has an associated trion family ($X^C$ and $X_z^C$, respectively) that inherits the selection rules (see Fig. 4c),[73] whose recombination results in a broad PL-band at the energy of the peak around 1.640 eV, as shown in Fig. 1c, due to the coupling of light to both to IP and OP dipoles. These facts arise trions as the main source of OP light able to excite WGMs of dielectric microspheres.

### 3. Conclusions

To summarize, we have demonstrated a spectroscopic method that enables the identification of the out-of-plane component of radiating dipoles. This would allow to identify, at any temperature, potential sources for radially polarized light in monolayer transition-metal dichalcogenides, which may have applications in imaging, phase modulation, and diffractive optics, due to the cylindrical symmetry of their light. Also, this method opens up new routes for the realization of planar devices efficiently exploiting the fundamental properties of atomically thin materials. The technique does not require magnetic fields and can be easily implemented, making it accessible, versatile, and technologically relevant. To evidence the effectiveness of this technique, we have performed a spectroscopical analysis of the luminescent signal of InSe nanosheets and monolayers of WSe$_2$. This study evidences that excitonic complexes with binding energies larger than that of the room-temperature free-excitons by even 150 meV (such as trions) have an important out-of-plane dipolar orientation, which makes them suitable candidates for their incorporation into planar devices with potential applications for information processing and on-chip communications. Interestingly, the wide energy range observed here for optical transitions in monolayer WSe$_2$ at room temperature with an important out-of-plane dipolar component also corresponds to the region where quantum emitters are usually observed at low temperature. Results reported here suggest that quantum light from these emitters are expected to efficiently couple into photonic chips to develop optimal quantum devices based on two-dimensional semiconductors.

4. **Materials and methods**

InSe and WSe$_2$ 2D nanosheets have been micromechanically exfoliated using the well-known scotch-tape technique. InSe monocrystals used here to prepare the nanosheets were cleaved perpendicular to the (001) direction from an ingot grown by the Bridgman method from a nonstoichiometric In$_{1.05}$Se$_{0.95}$ melt. To act as n-dopant, tin, in a content 0.01%, was introduced previously to growth. From these ingots, thin n-doped InSe samples were cleaved and used to prepare atomically thin InSe nanosheets. InSe samples have been then directly transferred onto Si substrates coated with 285 nm of SiO$_2$, previously cleaned with acetone, ethanol and isopropanol in an ultrasound bath.

p-doped bulk WSe$_2$ (from HQgraphene) has been used here to obtain WSe$_2$ nanosheets, which were obtained by a micromechanical exfoliation method to a polydimethylsiloxane (PDMS) stamp. Monolayers of WSe$_2$ were distinguished from thicker ones by means of their PL emission at room temperature and then transferred onto the Si/SiO$_2$ (285nm) cleaned substrates through the all-dry viscoelastic technique. The samples have been identified via optical contrast using a Zeiss Axio Scope.a1 microscope with an Axiocam ERc 5s camera.

SiO$_2$ microspheres solved in H$_2$O (from Sigma-Aldrich) have been used, with a nominal diameter of 5μm (4.83 ± 0.19 μm). To have a sparse concentration of these microspheres in solution, different aliquots in ethanol were prepared, and dropped onto the exfoliated substrates via spin coating, after which high vacuum is applied to force evaporation and avoid H$_2$O residues. In the case of ML WSe$_2$ nanosheets, microspheres in their surroundings were shoved until precisely placed on top of the nanosheets, by using a tip probe attached to the transfer setup micromanipulators.

Micro-PL measurements have been performed in a Horiba Scientific Xplora micro-Raman system using a 532nm CW excitation laser, not exceeding 70μW of power in 7s acquisition time measurements in InSe and 10μW in 1s acquisition time measurements in WSe$_2$ to prevent overheating. The optical excitation and collection spots are typically around 1 μm$^2$.

In order to calculate numerically the WGMs of the spherical micro-resonator, we used the COMSOL Multiphysics modeling software based on the FEM. A unit point dipole pointing to the center of the microsphere (OP dipole) or oriented perpendicularly (IP dipole) was introduced in our frequency-domain model through the RF module. To block troublesome reflections from outer boundaries of the computational domain, a perfectly matched layer of one-wavelength thickness surrounding the photonic structure was implemented. The power emitted by the electric point dipole in the presence of the sphere

($P$) and isolated dipole ($P_0$) were calculated by considering the power flux through a small-scale closed surface surrounding the emitter.

The optical absorption of monolayer WSe$_2$ has been computed within the framework of the Bethe-Salpeter Equation, as implemented in Yambo code.[74] The input of the BSE are the electronic states of WSe$_2$, calculated using density functional theory within the local-density approximation with the code Quantum Espresso.[75] We use fully relativistic pseudopotentials with semicore electrons for W. The simulations of the excitonic states have been performed in a 15x15x1 k-grid including 2 valence and 2 conduction bands. The vacuum distance between two periodic images is 30 Bohr.

## AUTHOR INFORMATION

**Corresponding Author**

\* Juan.F.Sanchez@uv.es

† Current Address: Institute of Photonics and Quantum Sciences, SUPA, Heriot-Watt University, Edinburgh EH14 4AS, UK.

**Author Contributions**

J.P.M.P. and J.F.S.R. conceived the study. D.A.P. performed experiments and analysis of experimental data. M.K. and C.J.Z.R. performed computational analyses and simulations on whispering gallery modes. A.M.S. performed ab initio computational analyses and simulations and contributed to theoretical descriptions. D.A.P, and J.F.S.R. wrote the manuscript, with extensive inputs from A.M.S., C.J.Z.R. and the rest of authors.

ACKNOWLEDGMENT

This work was made possible by the Horizon 2020 research and innovation program through the S2QUIP project (grant agreement No. 8204023) and by the Spanish MINECO through project No. TEC2017-86102-C2-1-R. D.A.-P acknowledges fellowship no. UV-INV-PREDOC17F1-539274 under the program "Atracció de Talent, VLCCAMPUS" of the University of Valencia. A. M.-S. acknowledges the Ramón y Cajal programme (grant RYC2018-024024-I; MINECO, Spain). Ab initio simulations were performed performed on the Tirant III cluster of the Servei d'Informática of the University of Valencia and on Mare Nostrum cluster of the Barcelona Supercomputing Center (project FI-2020-2-0033).

DATA AVAILABILITY

The data that support the findings of this study are available from the corresponding authors upon reasonable request. Correspondence and requests for materials should be addressed to J.F.S.R.

REFERENCES

1.  Won, R. Ultrafast nanoprobing. *Nat. Photonics* **4**, 882 (2010).

2.  Bae, S. H. *et al.* Integration of bulk materials with two-dimensional materials for physical coupling and applications. *Nat. Mater.* **18**, 550–560 (2019).

3.  Jia, Z. *et al.* Charge-Transfer-Induced Photoluminescence Properties of $WSe_2$ Monolayer-Bilayer Homojunction. *ACS Appl. Mater. Interfaces* **11**, 20566–20573 (2019).

4.  Flöry, N. *et al.* Waveguide-integrated van der Waals heterostructure photodetector at telecom

wavelengths with high speed and high responsivity. *Nat. Nanotechnol.* **15**, 118–124 (2020).

5. Kleemann, M. E. *et al.* Strong-coupling of WSe$_2$ in ultra-compact plasmonic nanocavities at room temperature. *Nat. Commun.* **8**, 1296 (2017).

6. Wang, H. *et al.* Plasmonically enabled two-dimensional material-based optoelectronic devices. *Nanoscale* **12**, 8095–8108 (2020).

7. Brotons-Gisbert, M., Martínez-Pastor, J. P., Ballesteros, G. C., Gerardot, B. D. & Sánchez-Royo, J. F. Engineering light emission of two-dimensional materials in both the weak and strong coupling regimes. *Nanophotonics* **7**, 253-267 (2018).

8. Reed, J. C. *et al.* Photothermal characterization of MoS$_2$ emission coupled to a microdisk cavity. *Appl. Phys. Lett.* **109**, 193109 (2016).

9. Ren, T., Song, P., Chen, J. & Loh, K. P. Whisper Gallery Modes in Monolayer Tungsten Disulfide-Hexagonal Boron Nitride Optical Cavity. *ACS Photonics* **5**, 353–358 (2018).

10. Mi, Y. *et al.* Tuning Excitonic Properties of Monolayer MoS$_2$ with Microsphere Cavity by High-Throughput Chemical Vapor Deposition Method. *Small* **13**, 1701694 (2017).

11. Salehzadeh, O., Djavid, M., Tran, N. H., Shih, I. & Mi, Z. Optically Pumped Two-Dimensional MoS2 Lasers Operating at Room-Temperature. *Nano Lett.* **15**, 5302–5306 (2015).

12. Zhao, L. *et al.* High-Temperature Continuous-Wave Pumped Lasing from Large-Area Monolayer Semiconductors Grown by Chemical Vapor Deposition. *ACS Nano* **12**, 9390–9396 (2018).

13. Schuller, J. A. *et al.* Orientation of luminescent excitons in layered nanomaterials. *Nat. Nanotechnol.* **8**, 271–276 (2013).

14. Brotons-Gisbert, M. *et al.* Out-of-plane orientation of luminescent excitons in atomically thin indium selenide flakes. *Nat. Commun.* 10, 3913 (2019).

15. Chernikov, A. *et al.* Exciton binding energy and nonhydrogenic Rydberg series in monolayer WS$_2$. *Phys. Rev. Lett.* **113**, 076802 (2014).

16. Borghardt, S. *et al.* Radially polarized light beams from spin-forbidden dark excitons and trions in

monolayer WSe$_2$. *Opt. Mater. Express* **10**, 1273–1285 (2020).

17. Mu, X., Wu, S., Cheng, L. & Fu, H. Y. Edge couplers in silicon photonic integrated circuits: A review. *Appl. Sci.* **10**, 1538 (2020).

18. Ren, T. & Loh, K. P. On-chip integrated photonic circuits based on two-dimensional materials and hexagonal boron nitride as the optical confinement layer. *J. Appl. Phys.* **125**, 230901 (2019).

19. Wang, Z., Zervas, M. N., Bartlett, P. N. & Wilkinson, J. S. Surface and waveguide collection of Raman emission in waveguide-enhanced Raman spectroscopy. *Opt. Lett.* **41**, 4146–4149 (2016).

20. Jun, Y. C., Briggs, R. M., Atwater, H. A. & Brongersma, M. L. Broadband enhancement of light emission in silicon slot waveguides. *Opt. Express* **17**, 7479 (2009).

21. Tan, R. & Huang, H. Spontaneous-emission coupling from an excited atom into a symmetrical metal-cladding optical waveguide. *Chinese Phys. Lett.* **31**, 084205 (2014).

22. Verhart, N. R., Lepert, G., Billing, A. L., Hwang, J. & Hinds, E. A. Single dipole evanescently coupled to a multimode waveguide. *Opt. Express* **22**, 19633 (2014).

23. Davanco, M. I. & Srinivasan, K. Efficient spectroscopy of single embedded emitters using optical fiber taper waveguides. *Opt. Express* **17**, 10542 (2009).

24. Liebermeister, L. *et al.* Tapered fiber coupling of single photons emitted by a deterministically positioned single nitrogen vacancy center. *Appl. Phys. Lett.* **104**, 031101 (2014).

25. Tamalampudi, S. R. *et al.* High performance and bendable few-layered InSe photodetectors with broad spectral response. *Nano Lett.* **14**, 2800–2806 (2014).

26. Mudd, G. W. *et al.* High Broad-Band Photoresponsivity of Mechanically Formed InSe-Graphene van der Waals Heterostructures. *Adv. Mater.* **27**, 3760–3766 (2015).

27. Lei, S. *et al.* An atomically layered InSe avalanche photodetector. *Nano Lett.* **15**, 3048–3055 (2015).

28. Mudd, G. W. *et al.* Tuning the bandgap of exfoliated InSe nanosheets by quantum confinement. *Adv. Mater.* (2013) doi:10.1002/adma.201302616.

29. Sánchez-Royo, J. F. *et al.* Electronic structure, optical properties, and lattice dynamics in atomically thin indium selenide flakes. *Nano Res.* **7**, 1556–1568 (2014).

30. Brotons-Gisbert, M. *et al.* Nanotexturing To Enhance Photoluminescent Response of Atomically Thin Indium Selenide with Highly Tunable Band Gap. *Nano Lett.* **16**, 3221–3229 (2016).

31. Bandurin, D. A. *et al.* High electron mobility, quantum Hall effect and anomalous optical response in atomically thin InSe. *Nat. Nanotechnol.* **12**, 223–227 (2017).

32. Hu, F. *et al.* Imaging exciton-polariton transport in $MoSe_2$ waveguides. *Nat. Photonics* **11**, 356–360 (2017).

33. Autere, A. *et al.* Nonlinear Optics with 2D Layered Materials. *Adv. Mater.* **30**, 1705963 (2018).

34. Zhang, L., Gogna, R., Burg, W., Tutuc, E. & Deng, H. Photonic-crystal exciton-polaritons in monolayer semiconductors. *Nat. Commun.* **9**, 713 (2018).

35. Elshaari, A. W., Pernice, W., Srinivasan, K., Benson, O. & Zwiller, V. Hybrid integrated quantum photonic circuits. *Nat. Photonics* **14**, 285–298 (2020).

36. Gonzalez Marin, J. F., Unuchek, D., Watanabe, K., Taniguchi, T. & Kis, A. $MoS_2$ photodetectors integrated with photonic circuits. *npj 2D Mater. Appl.* **3**, 14 (2019).

37. Thakar, K. & Lodha, S. Optoelectronic and photonic devices based on transition metal dichalcogenides. *Mater. Res. Express* **7**, 14002 (2019).

38. Wang, G. *et al.* In-Plane Propagation of Light in Transition Metal Dichalcogenide Monolayers: Optical Selection Rules. *Phys. Rev. Lett.* **119**, 047401 (2017).

39. Tang, Y., Mak, K. F. & Shan, J. Long valley lifetime of dark excitons in single-layer $WSe_2$. *Nat. Commun.* **10**, 4047 (2019).

40. Shi, J. *et al.* Enhanced Trion Emission and Carrier Dynamics in Monolayer $WS_2$ Coupled with Plasmonic Nanocavity. *Adv. Opt. Mater.* **8**, 2001147 (2020).

41. Mak, K. F. *et al.* Tightly bound trions in monolayer $MoS_2$. *Nat. Mater.* **12**, 207–211 (2013).

42. Ross, J. S. *et al.* Electrical control of neutral and charged excitons in a monolayer semiconductor. *Nat. Commun.* **4**, 1474 (2013).

43. Liu, E. *et al.* Gate Tunable Dark Trions in Monolayer WSe$_2$. *Phys. Rev. Lett.* **123**, 027401 (2019).

44. Singh, A. *et al.* Long-Lived Valley Polarization of Intravalley Trions in Monolayer WSe$_2$. *Phys. Rev. Lett.* **117**, 257402 (2016).

45. Li, Z. *et al.* Revealing the biexciton and trion-exciton complexes in BN encapsulated WSe$_2$. *Nat. Commun.* **9**, 3719 (2018).

46. Huang, J., Hoang, T. B. & Mikkelsen, M. H. Probing the origin of excitonic states in monolayer WSe$_2$. *Sci. Rep.* **6**, 22414 (2016).

47. Jadczak, J. *et al.* Probing of free and localized excitons and trions in atomically thin WSe$_2$, WS$_2$, MoSe$_2$ and MoS$_2$ in photoluminescence and reflectivity experiments. *Nanotechnology* **28**, 395702 (2017).

48. Harats, M. G., Kirchhof, J. N., Qiao, M., Greben, K. & Bolotin, K. I. Dynamics and efficient conversion of excitons to trions in non-uniformly strained monolayer WS$_2$. *Nat. Photonics* **14**, 324–329 (2020).

49. Brotons-Gisbert, M. *et al.* Optical and electronic properties of 2H−MoS$_2$ under pressure: Revealing the spin-polarized nature of bulk electronic bands. *Phys. Rev. Mater.* **2**, 054602 (2018).

50. Wang, Z. *et al.* Excitonic complexes and optical gain in two-dimensional molybdenum ditelluride well below the Mott transition. *Light Sci. Appl.* **9**, 2047–7538 (2020).

51. Courtade, E. *et al.* Charged excitons in monolayer WSe$_2$: Experiment and theory. *Phys. Rev. B* **96**, 085302 (2017).

52. Branny, A., Kumar, S., Proux, R. & Gerardot, B. D. Deterministic strain-induced arrays of quantum emitters in a two-dimensional semiconductor. *Nat. Commun.* **8**, 15053 (2017).

53. Brotons-Gisbert, M. *et al.* Coulomb blockade in an atomically thin quantum dot coupled to a tunable Fermi reservoir. *Nat. Nanotechnol.* **14**, 442–446 (2019).

54. Chakraborty, C., Vamivakas, N. & Englund, D. Advances in quantum light emission from 2D materials. *Nanophotonics* **8**, 2017–2032 (2019).

55. Molas, M. R. *et al.* Probing and Manipulating Valley Coherence of Dark Excitons in Monolayer WSe$_2$. *Phys. Rev. Lett.* **123**, 096803 (2019).

56. Gao, Y., Weidman, M. C. & Tisdale, W. A. CdSe Nanoplatelet Films with Controlled Orientation of their Transition Dipole Moment. *Nano Lett.* **17**, 3837–3843 (2017).

57. Zhou, Y. *et al.* Probing dark excitons in atomically thin semiconductors via near-field coupling to surface plasmon polaritons. *Nat. Nanotechnol.* **12**, 856–860 (2017).

58. Chen, S. Y., Goldstein, T., Taniguchi, T., Watanabe, K. & Yan, J. Coulomb-bound four- and five-particle intervalley states in an atomically-thin semiconductor. *Nat. Commun.* **9**, 3717 (2018).

59. Liu, Y., Li, H., Qiu, C., Hu, X. & Liu, D. Layer-dependent signatures for exciton dynamics in monolayer and multilayer WSe2 revealed by fluorescence lifetime imaging measurement. *Nano Res.* **13**, 661–666 (2020).

60. Kang, W. T. *et al.* Direct growth of doping controlled monolayer WSe$_2$ by selenium-phosphorus substitution. *Nanoscale* **10**, 11397–11402 (2018).

61. Li, P.-Y. *et al.* Unusual imaging properties of superresolution microspheres. *Opt. Express* **24**, 16479 (2016).

62. Lu, D. & Liu, Z. Hyperlenses and metalenses for far-field super-resolution imaging. *Nat. Commun.* **3**, 1205 (2012).

63. Lee, S., Li, L., Ben-Aryeh, Y., Wang, Z. & Guo, W. Overcoming the diffraction limit induced by microsphere optical nanoscopy. *J. Optics* **15**, 125710 (2013).

64. Wang, G. *et al.* In-Plane Propagation of Light in Transition Metal Dichalcogenide Monolayers: Optical Selection Rules. *Phys. Rev. Lett.* **119**, 047401 (2017).

65. Asano, M. *et al.* Observation of optomechanical coupling in a microbottle resonator. *Laser Photonics Rev.* **10**, 603–611 (2016).

66. Roselló-Mechó, X. *et al.* Parametrical Optomechanical Oscillations in PhoXonic Whispering Gallery Mode Resonators. *Sci. Rep.* **9**, 7163 (2019).

67. Chiasera, A. *et al.* Spherical whispering-gallery-mode microresonators. *Laser Photonics Rev.* **4**, 457–482 (2010).

68. Chew, H. Transition rates of atoms near spherical surfaces. *J. Chem. Phys.* **87**, 1355–1360 (1987).

69. Stratton, J. A. *Electromagnetic Theory*. *Electromagnetic Theory* (John Wiley & Sons, Inc., (2015), 2015). doi:10.1002/9781119134640.

70. Ruppin, R. Decay of an excited molecule near a small metal sphere. *J. Chem. Phys.* **76**, 1681–1684 (1982).

71. Molina-Sánchez, A., Sangalli, D., Wirtz, L. & Marini, A. Ab Initio Calculations of Ultrashort Carrier Dynamics in Two-Dimensional Materials: Valley Depolarization in Single-Layer WSe2. *Nano Lett.* **17**, 4549–4555 (2017).

72. Brotons-Gisbert, M. *et al.* Spin–layer locking of interlayer excitons trapped in moiré potentials. *Nat. Mater.* **19**, 630–636 (2020).

73. Drüppel, M., Deilmann, T., Krüger, P. & Rohlfing, M. Diversity of trion states and substrate effects in the optical properties of an MoS2 monolayer. *Nat. Commun.* **8**, 2117 (2017).

74. Sangalli, D. *et al.* Many-body perturbation theory calculations using the yambo code. *J. Phys. Condens. Matter* **31**, 325902 (2019).

75. Giannozzi, P. *et al.* QUANTUM ESPRESSO: A modular and open-source software project for quantum simulations of materials. *J. Phys. Condens. Matter* **21**, 395502 (2009).

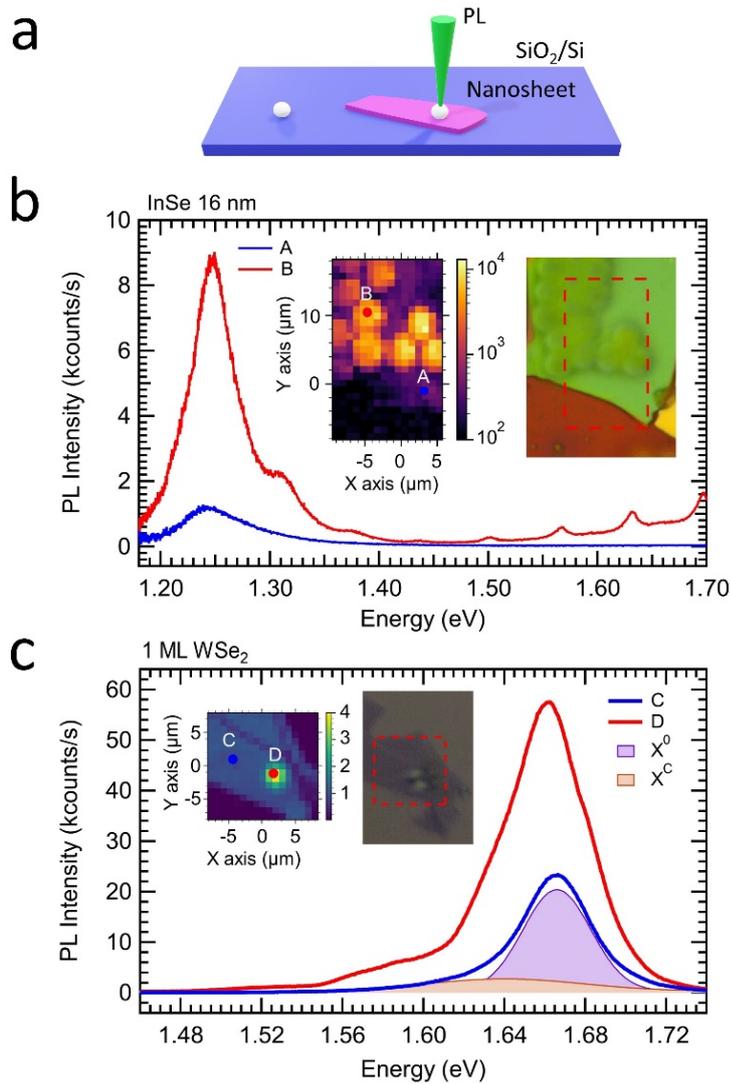

**Figure 1. Photoluminescence response of 2D nanosheets partly covered with SiO₂ microspheres**. (a) Illustration of the measurement process of the PL response of nanosheets partly covered by dielectric microspheres. (b) Micro-PL spectra acquired in two points of the 16 nm thick InSe nanosheet deposited on a SiO$_2$/Si substrate, which is shown in the optical image in the inset. A certain number of microspheres can be observed in the optical image, as blurred spheres since the image was taking focusing on the nanosheet. The inset includes a PL map acquired in a region of the nanosheet delimited by the red dashed rectangle depicted on the optical image. The PL-signal integrated intensity of each point of the PL map follows the color-code indicated on the right side of the PL map, revealing a clear PL intensity enhancement in the region of the nanosheet covered with microspheres. The two μ-PL spectra shown in the main graph were acquired in each one of the two points marked in the PL map, one in a bare region of the nanosheet (A point) and the other in a point covered by a microsphere (B point). (c) Micro-PL spectra acquired in two points of a ML WSe$_2$ nanosheet deposited on a SiO$_2$/Si substrate, which is shown in the optical image in the inset. In this case, a microsphere was purposely pushed on top of the WSe$_2$ nanosheet, which appears on the optical image as a burred bump. The inset includes a PL map acquired in a region of the nanosheet delimited by the red dashed rectangle depicted on the optical image. As before, the PL map reveals a clear PL intensity enhancement in the region of the nanosheet covered with the microsphere. The two μ-PL spectra shown in the main graph were acquired in each one of the two points marked in the PL map, one in a bare region of the nanosheet (C point) and the other on a point covered by a microsphere (D point). The μ-PL spectrum measured in the C point was deconvoluted into two gaussian components (shown on the graph), one attributed to the neutral exciton $X^0$ and the other to charged excitons $X^c$.

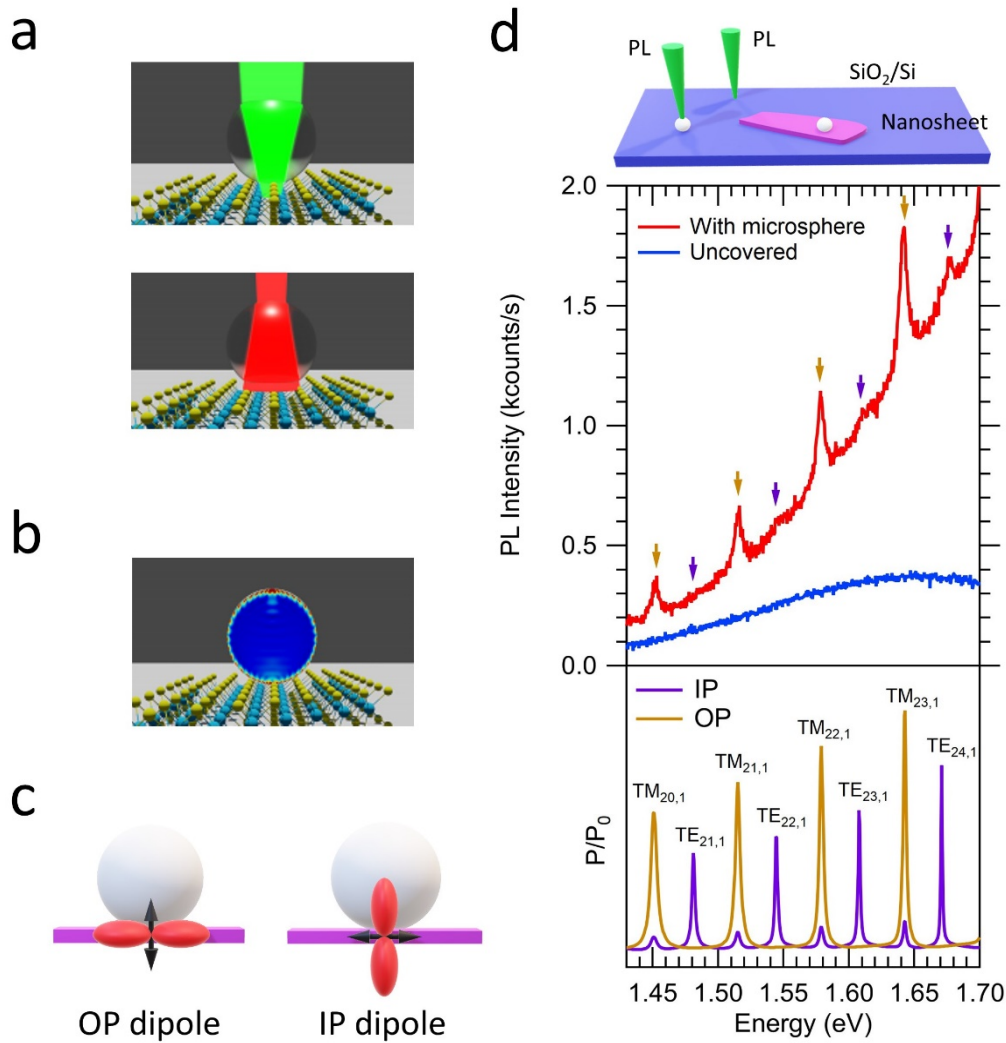

**Figure 2. Optical properties of whispering gallery modes in dielectric microspheres.** (a) Illustration of the focusing effects of the incident (top panel) and emitted light (bottom panel) produced when a dielectric microsphere is placed on top of an emitting layer. (b) Illustration of an eventual WGM excited by the coupling of emitted light with the modes of the microspherical resonator. (c) Illustration of the two possible orientations of radiative dipoles that may excite WGMs in the dielectric microspheres on top. Red lobes illustrate dipolar far-field intensity in free space. (d) Micro-PL spectra measured in an amorphous $SiO_2$/Si substrate which contains native radiative centres spread within the $SiO_2$ overlayer. The µ-PL spectra of these centres have been recorded on points of the $SiO_2$/Si substrate with and without dielectric microspheres on top (see illustration depicted on the plot). Comparing these two µ-PL spectra, only the one measured in the point covered with a microsphere shows the presence of quasi-equally spaced peaks due to the coupling of the light emitted excitation of WGMs of the microsphere. The lower panel of the plot shows the spectral response response of the normalized power calculated for a dipolar emission coupled to the WGMs of a spherical dielectric microresonator of index of refraction 1.45 and 4.66 µm in diameter, when these emitting dipoles are IP (purple curve) or OP (orange curve) oriented, as it has been illustrated in (c). The TM or TE character of each WGMs has been indicated on each resonance peak, as referred to the radial component of the vector field.

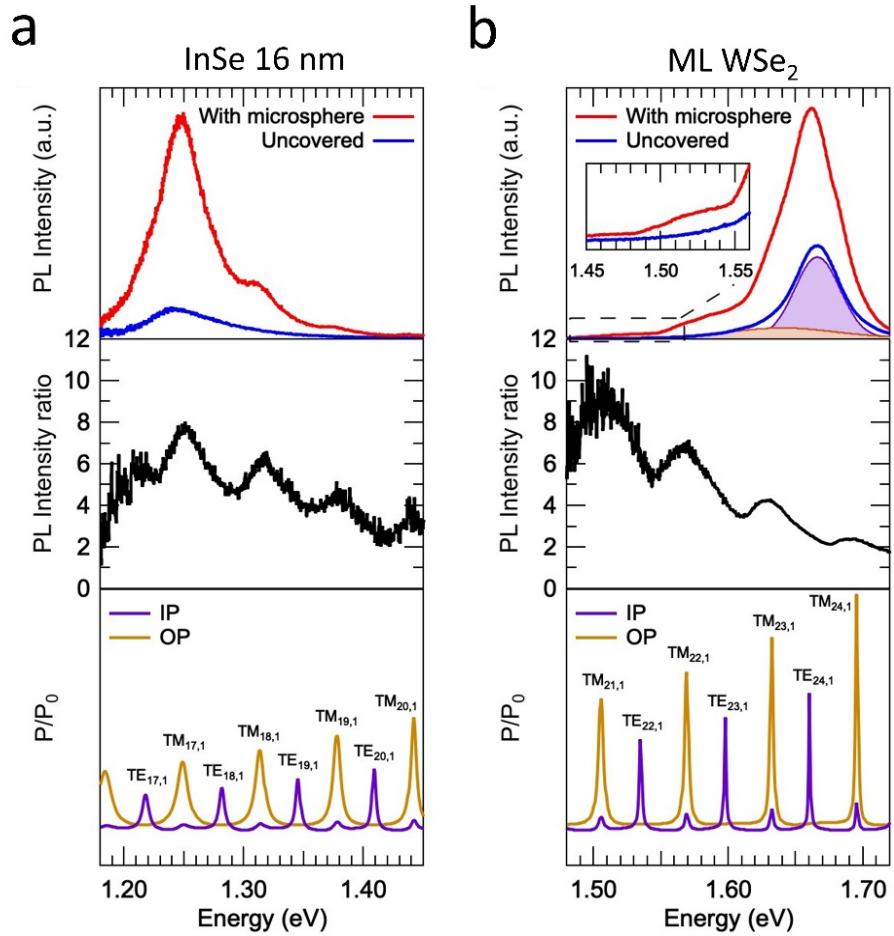

**Figure 3. Revealing the OP orientation of dipoles by the excitation of WGMs of dielectric microspheres.** Top panels: μ-PL spectra acquired in the (a) 16 nm thick InSe nanosheet and (b) ML WSe$_2$ which have been already shown in Figure 1b and 1c, respectively. Middle panels: Intensity ratio between the PL spectra shown in their respective top plots, which evidence the coupling of emitted light to the WGMs of dielectric microspheres. Bottom panels: Spectral response calculated for a dipolar emission coupled to the WGMs of a spherical microresonator of 4.69 μm in diameter, when these emitting dipoles are IP (purple curve) or OP (orange curve) oriented. The TM or TE character of each WGMs has been indicated on each resonance peak.

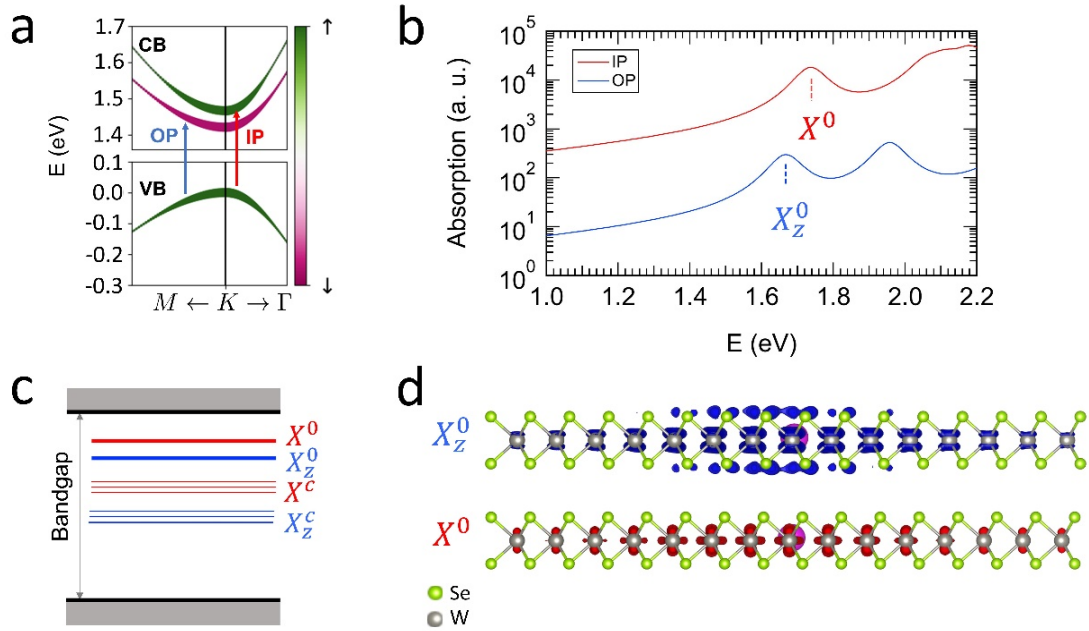

**Figure 4. Optical selection rules for excitons and trions.** (a) Band structure of monolayer WSe$_2$ calculated by using LDA along the M-K-Γ high-symmetry direction. The color code indicates the spin projection of each band. The width indicates the contribution of each band, at the K-point, to the excitonic wavefunction of the excitons X$^0$ (IP transition) and X$_z^0$ (OP transition). (b) Optical absorption for IP and OP configurations, in logarithmic scale. The ground excitonic states of each spectra are denoted as X$^0$ and X$^0_z$. (c) Scheme of energy levels of excitons and trions. The trions X$^c$ and X$^c_z$ inherits the selection rules of the excitons X$^0$ and X$^0_z$, respectively. (d) Wave functions of the excitons X$^0$ and X$^0_z$. The position of the hole is fixed (pink sphere) near the W atom and we represent the electronic density. The maximum is set to 1 and we fix the isosurface value to 0.1 for both wave functions.

# Supplementary Information

# Out-of-plane emission of trions in monolayer WSe$_2$ revealed by whispering gallery modes of dielectric microresonators

*Daniel Andres-Penares*[1,†]*, Mojtaba Karimi Habil*[2,3]*, Alejandro Molina-Sánchez*[1]*, Carlos J. Zapata-Rodríguez*[2]*, Juan P. Martínez-Pastor*[1,4]*, and Juan F. Sánchez-Royo*[1,4]

[1] ICMUV, Instituto de Ciencia de Materiales, Universidad de Valencia, P.O. Box 22085, 46071 Valencia, Spain

[2] Department of Optics and Optometry and Vision Sciences, University of Valencia, c/ Dr. Moliner 50, Burjassot 46100, Spain

[3] Faculty of Physics, University of Tabriz, 51664 Tabriz, Iran

[4] MATINÉE: CSIC Associated Unit-(ICMM-ICMUV of the University of Valencia), Universidad de Valencia, P.O. Box 22085, 46071 Valencia, Spain

# 1 Spherical vector wave functions expansion

In a free-of-sources region ($r > R$), a well-behaved electromagnetic field may be represented in terms of the spherical, vector wave functions [S1],

$$\mathbf{E} = \sum_{n=1}^{+\infty} \sum_{m=-n}^{n} (a_{nm}\mathbf{M}_{nm}^{(3)} + b_{nm}\mathbf{N}_{nm}^{(3)}), \tag{1a}$$

$$\mathbf{H} = -\frac{ik}{\omega\mu\mu_0} \sum_{mn} (a_{nm}\mathbf{N}_{nm}^{(3)} + b_{nm}\mathbf{M}_{nm}^{(3)}), \tag{1b}$$

where the term $n = 0$ is excluded for a non-static field (note that $P_0^0(\cos\theta) = 1$ and therefore $Y_0^0 = 1/2\sqrt{\pi}$).

## 1.1 Classification of wave fields

It will be recalled that the radial component of every function $\mathbf{M}_{nm}^{(3)}$ is zero, that is $\mathbf{M}_{nm}^{(3)} \cdot \hat{r} = 0$. Hence:

- **TM$_r$ waves.** If the coefficients $a_{mn}$ are all zero, only the $b_{mn}$ being excited, the field has a radial component of $\mathbf{E}$ but the magnetic vector $\mathbf{H}$ is always perpendicular to the radius vector $\mathbf{r} = r\hat{r}$. The oscillations whose amplitudes are represented by the coeficients $b_{mn}$ are referred to as *transverse magnetic* (TM).

- **TE$_r$ waves.** If only the $a_{mn}$ are excited, these oscillations may also be said to be *transverse electric* (TE).

## 1.2 Useful expansions: electric point dipole

The vector spherical harmonics can be interpreted as multipole fields [S2]. We will only illustrate here this important aspect on a simple example. Let us for example consider [S3, Appendix H]

$$\mathbf{M}_{1,0}^{(3)}(r, \theta, \phi) = \exp(ikr)\left[-\frac{1}{kr} - \frac{i}{(kr)^2}\right]\sqrt{\frac{3}{4\pi}}\sin\theta\hat{\phi}, \tag{2a}$$

$$\mathbf{N}_{1,0}^{(3)}(r, \theta, \phi) = \exp(ikr)\left[-\frac{1}{(kr)^2} - \frac{i}{(kr)^3}\right]\sqrt{\frac{3}{\pi}}\cos\theta\hat{r} \tag{2b}$$

$$-\frac{1}{kr}\exp(ikr)\left[-i + \frac{1}{kr} + \frac{i}{(kr)^2}\right]\sqrt{\frac{3}{4\pi}}\sin\theta\hat{\theta}, \tag{2c}$$

One may recognize in the latter expression the electric field created by an electric dipole at the origin and oriented along $OZ$: $\mathbf{p} = p\hat{z}$. More precisely:

1. The electric field and the magnetic field of such an electric dipole set at the origin and evaluated in $r > 0$ are:

$$\mathbf{E} = \frac{ik^3 p}{\sqrt{12\pi\epsilon_0\epsilon}} \mathbf{N}^{(3)}_{1,0}, \quad (3a)$$

$$\mathbf{H} = \frac{\omega k^2 p}{\sqrt{12\pi}} \mathbf{M}^{(3)}_{1,0}. \quad (3b)$$

That is, only the coefficient $b^{(3)}_{10}$ is different from zero.

2. The electric field of an electric dipole of moment $\mathbf{p} = p\hat{z}$ set along the polar $z$-axis at a distance $r_p > 0$, when evaluated at points in the region $r > r_p$, can be set as [S4]

$$\mathbf{E} = \frac{ik^2 p}{\sqrt{4\pi}r_p\epsilon_0\epsilon} \sum_{n=1}^{\infty} \sqrt{2n+1}\, j_n(kr_p)\mathbf{N}^{(3)}_{n,0}. \quad (4)$$

In this case, the coefficients $b_{n0}$ are different from zero.

3. The spherical vector wave functions $\mathbf{N}^{(3)}_{1,\pm 1}$ are used to represent the electric field of an electric dipole set at the origin and oriented perpendicular to the $z$-axis through the non-zero coefficients $b_{1\pm 1}$ [S3].

4. When the electric dipole is displaced along the polar axis, and again is oriented perpendicular to this axis, the electric field can be set in terms of spherical vector wave functions associated with non-vanishing coefficients $a_{n\pm 1}$ and $b_{n\pm 1}$ [S4].

5. In the presence of a multilayered spherical scatterer with center at origin, the scattered signal can be set as a linear combination of the same set of spherical vector wave functions contributing to characterize the excited field of the dipole point source.

## 2 Effects of the microsphere parameters on the WGMs observed in 2D nanosheets

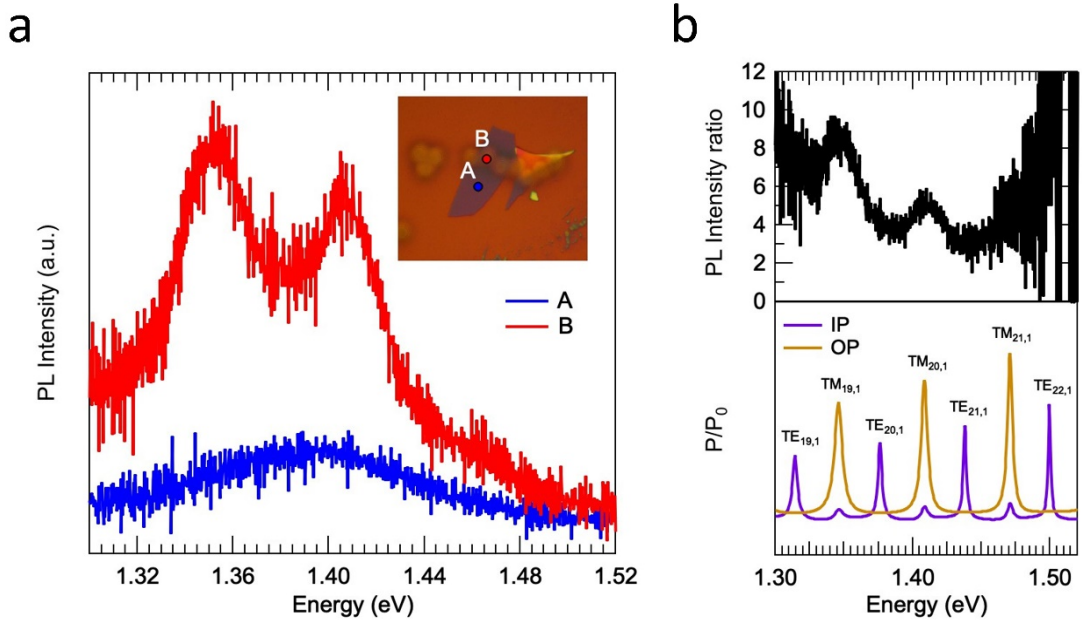

**Figure S1. WGMs in 2D InSe nanosheets.** (a) Micro-PL spectra acquired in two points of a 6.5 nm thick InSe nanosheet deposited on a SiO$_2$/Si substrate, which is shown in the optical image in the inset as a dark nanosheet [S5]. The A and B μ-PL spectra were acquired in bare region of the nanosheet and in a point covered by a microsphere, which correspond to the positions labelled on the optical image, respectively. (b) top plot: Intensity ratio between the PL spectra shown in (a). Bottom plots: Spectral response calculated for a dipolar emission coupled to the WGMs of a spherical microresonator of a diameter of 4.80 μm, when these emitting dipoles are IP (purple curve) or OP (orange curve) oriented. The TM or TE character of each WGMs has been indicated on each resonance peak.

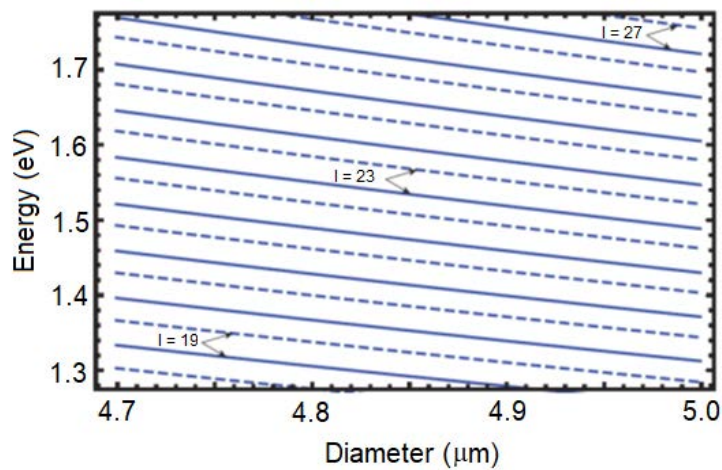

**Figure S2. Diameter dependence of the WGMs resonances.** Energy at which WGMs are expected to appear, as a function of the diameter of the SiO$_2$ microsphere. For the calculations, we have assumed that the microspheres, with a refraction index of 1.4607, are immersed in air. Solid lines correspond to the $TE_{l,1}$ modes whereas dashed lines correspond to $TM_{l,1}$ modes. The $l$-order of the TE and TM modes has been indicated for some of the curves.

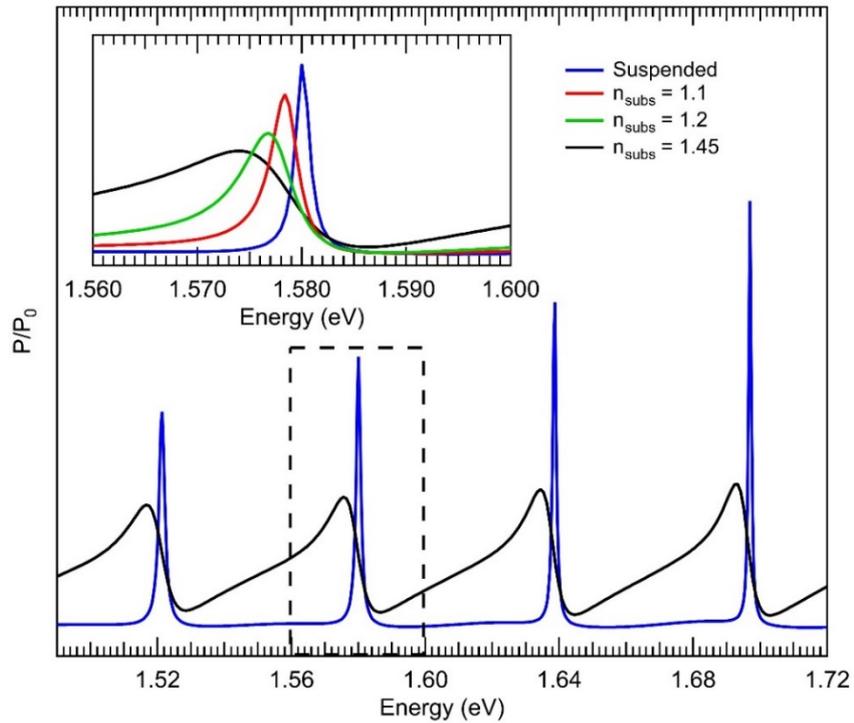

**Figure S3. Substrate effects on the WGMs line shape.** Influence of the substrate on the line shape of the WGMs. For the calculations, we have assumed that the dipole is located at the vicinity of the microsphere and that the substrate has a refractive index ranging from 1.0 (suspended microspheres) to 1.45 ($SiO_2$ substrate). The inset shows a zoom of one these resonances.

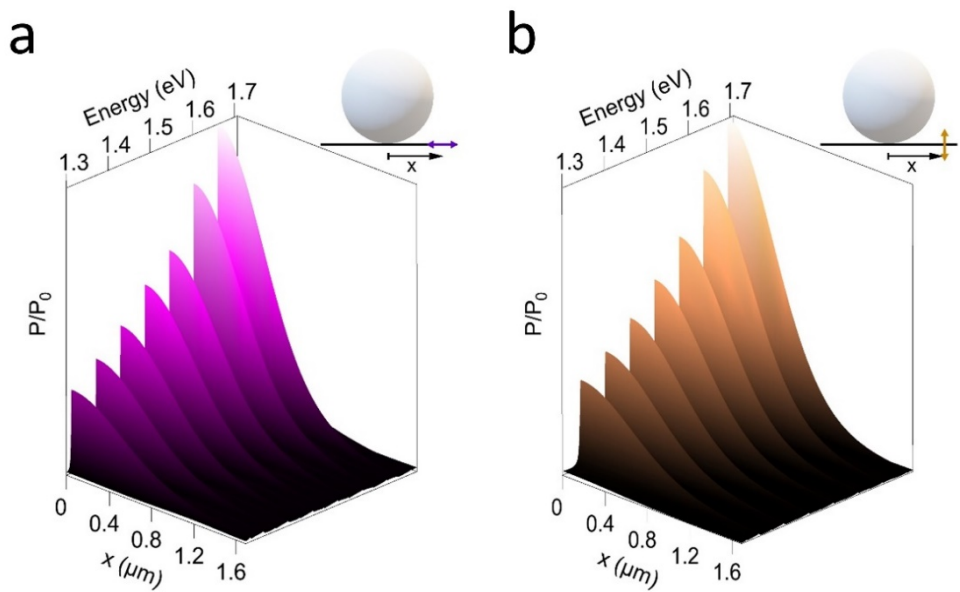

**Figure S4. Effective excitation of WGMs.** (a)-(b) Relative strength of the WGMs resonances originated by the coupling of the light emitted by a dipole located a distance x from the microsphere vertical symmetry axis, for the case of an IP and OP dipole, respectively. Note that both OP and IP modes attenuate a factor 0.5 already for x=0.6 μm, a quarter of the microsphere radius.

# 3  Symmetry analysis of optical selection rules of monolayer WSe₂

The main properties of the photoluminescence spectra of monolayer WSe2 are determined by the optical selection rules at the K point. Figure S5 shows the bands at K, labelled with the symmetry representation. The up and down arrow indicates the dominant spin projection. At the opposite K point (K'), the spin of each band is reversed.

In absence of spin-orbit interaction, the optical selection rules only allow absorption for in-plane light. The out-of-plane transition is possible only due to the spin-orbit interaction and the resulting mixing of valence bands of opposite spin projection. We have also marked the magnitude of the spin-orbit splitting at the conduction and valence bands (Figure S5). It is a common trend in transition metal dichalcogenides a larger magnitude in the valence band [S6].

The OP transition at one of the two K-points involves the $\Gamma_{10}(\uparrow)$ from the CB and $\Gamma_8(\downarrow)$ from the VB, where $\Gamma_n$ are the irreducible representations, and the product is $\Gamma_{10}(\uparrow) \times \Gamma_8(\downarrow)$. The spin-orbit interaction induces a band mixing among the valence band $\Gamma_8(\downarrow)$ and deeper ones (not shown here). These deeper valence bands ($v'$) transform with the same representations but with reversed spin. Therefore, states at the top valence band can be expressed as $\alpha U_{\Gamma_8(\downarrow)} + \beta U_{v'(\uparrow)}$. The coefficient β depends on the strength of the spin-orbit interaction. Due to this mixture, z-polarized (i.e., OP) optical transitions become possible between the valence and conduction band with opposite spin, as the band $v'(\uparrow)$ has the same spin orientation than the conduction band $\Gamma_{10}$, and therefore the total angular momentum (including valley, spin, and orbital component) is strictly conserved. Nevertheless, the oscillator strength is much weaker than in the case of IP transitions. A derivation based in group theory is presented in the Supplementary Information of Ref. 64.

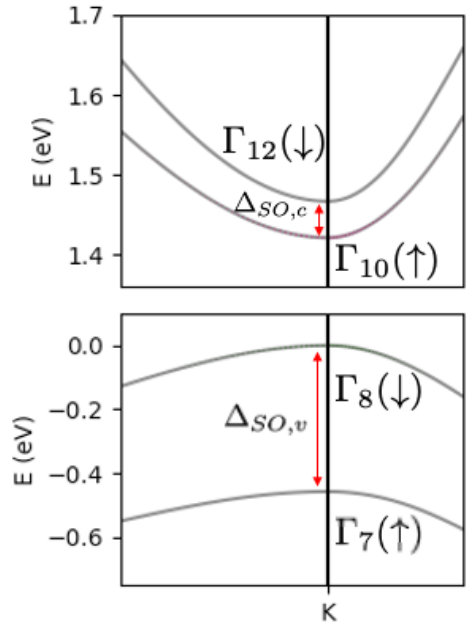

**Figure S5: Electronic structure at the K point of monolayer WSe₂.** Band dispersion along the MKΓ high symmetry directions of the lowest conduction bands (top panel) and higher valence bands (bottom panel). Labels indicate the symmetry representation of the bands at K point. Spin-orbit valence ($\Delta_{SO,v}$) and conduction band ($\Delta_{SO,c}$) splitting at Γ point has been indicated.

## Supplementary References

[S1] Stratton, J. A. Electromagnetic theory. McGraw-Hill College, 1941.
[S2] Jackson, J. D. Classical electrodynamics, third edition. John Wiley & Sons, 1999.
[S3] Le Ru E. & Etchegoin, P. Principles of surface-enhanced Raman spectroscopy and related plasmonic effects. Elsevier, 2009.
[S4] Enderlein, J. Electrodynamics of fluorescence, 2003. Technical report.
[S5] Brotons-Gisbert, M., Andres-Penares, D., Martínez-Pastor, J. F., Cros, A. & Sánchez-Royo, J. F. Optical contrast of 2D InSe on $SiO_2$/Si and transparent substrates using bandpass filters. *Nanotechnology* **28** 115706 (2017).
[S6] Molina-Sánchez, A., Hummer, K. & Wirtz, L. Vibrational and optical properties of $MoS_2$: From monolayer to bulk. *Surface Science Reports* **70**, 554-586 (2015).